\documentclass{osa-article}

\journal{oe}

\articletype{Research Article}

\begin{document}

\title{Deep-learning-assisted optical communication with discretized state space of structured light}

\author{Minyang Zhang,\authormark{1} Dong-Xu Chen,\authormark{2,3,*}  Pengxiang Ruan,\authormark{1} Jun Liu,\authormark{1} Jun-Long Zhao, \authormark{2} and Chui-Ping Yang\authormark{4,5}}

\address{\authormark{1}School of Science, Jiangsu University of Science and Technology, ZhenJiang 212003, China\\
\authormark{2}Quantum Information Research Center, Shangrao Normal University, Shangrao 334001, China\\
\authormark{3}Jiangxi Province Key Laboratory of Applied Optical Technology, Shangrao Normal University, Shangrao 334001, China\\
\authormark{4}School of Physics, Hangzhou Normal University, Hangzhou 311121, China\\
\authormark{5}yangcp@hznu.edu.cn}

\email{\authormark{*}chendx@sru.edu.cn} 



\begin{abstract}
The rich structure of transverse spatial modes of structured light has facilitated their extensive applications in quantum information and optical communication. The Laguerre-Gaussian (LG) modes, which carry a well-defined orbital angular momentum (OAM), consist of a complete orthogonal basis describing the transverse spatial modes of light. The application of OAM in free-space optical communication is restricted due to the experimentally limited OAM numbers and the complex OAM recognition methods.
Here, we present a novel method that uses the advanced deep learning technique for LG modes recognition. By discretizing the spatial modes of structured light, we turn the OAM state regression into classification.  A proof-of-principle experiment is also performed, showing that our method effectively categorizes OAM states with small training samples and high accuracy. By assigning each category a classical information, we further apply our approach to an image transmission task, demonstrating the ability to encode large data with low OAM number. This work opens up a new avenue for achieving high-capacity optical communication with low OAM number based on structured light.
\end{abstract}

\section{Introduction}
The orbital angular momentum (OAM) of light, characterized by its distinctive helical phase front $e^{i\ell\phi}$, has facilitated a pivotal breakthrough in the field of optics. The advent of OAM plays a crucial role in advancing the frontiers of many fields, such as optical communication, quantum information and quantum computation. The OAM of light  offers advantages over the well-studied spin angular momentum, i.e., the polarization of light, due to the inherent high dimensional peculiarity. For example, it has been shown that the OAM of light can be applied to realize high-dimensional quantum communication with larger alphabet and higher security \cite{PhysRevLett.88.127902,Chen_2017,PhysRevLett.108.143603,PhysRevApplied.11.064058,doi:10.1126/sciadv.1601915,Hu:20}. High-dimensional quantum entangled states can be generated with the OAM, which show larger violation of Bell nonlocality \cite{PhysRevLett.104.060401,Weiss_2016,PhysRevA.98.042105}. In the field of telecommunications, OAM is crucial for the development of high-capacity optical networks, especially in the enhancement of data transmission capabilities. Its unique ability to handle large data streams offers potential solutions to bandwidth limitations and efficiency challenges in fiber optical communication systems. Most importantly, as a cornerstone of high-dimensional quantum information processing, OAM has facilitated significant advancements in quantum computing and quantum encryption technology. The significance of OAM in these aspects is becoming ever more paramount, due to the increasing focus on higher bandwidth and denser information density nowadays. 

Previous works on the OAM of light focused on the generation and measurement of the OAM states. An OAM state with large $\ell$ and a simple method to recognize the OAM state are preferred in practical applications. However, the quantum numbers $\ell$ of experimentally generated OAM states are limited due to the imperfections of experimental devices. In 2016, Zeilinger's group generated the OAM states with quantum numbers up to 10,010 and realized quantum entanglement of these states \cite{10010}. On the other hand, the existing methods for measuring OAM states either cannot be applied in single photon level \cite{Sztul:06,PhysRevLett.101.100801,Guo:09,PhysRevLett.105.053904} or require complex measurement setups due to the high-dimension nature of OAM states \cite{PhysRevLett.88.257901,PhysRevLett.105.153601,CHEN20152530,PhysRevLett.119.180510,PhysRevLett.120.193904,Brandt:20,https://doi.org/10.1002/lpor.202300277}. Therefore, how to increase the channel capacity with limited OAM number is one active research area in OAM-based optical communication. Meanwhile, a flexible approach that can be adapted to the practical situations to recognize the OAM states is important for the applications of OAM.

Early applications of OAM in optical communication aimed to increase channel capacity through OAM multiplexing \cite{wang2012terabit,Terabit,yan2014high}. The OAM multiplexing technique treats OAM as one more degree of freedom in addition to other degrees of photons. Therefore, higher capacity could be achieved by exhausting available degrees of freedom of photons. However, only a subset of OAM states was used and the rich structure of the spatial modes was not fully harnessed. In recent years, the development of Machine Learning (ML) has promoted a new frontier in the research of photonic OAM. The enhanced capabilities of ML in data analysis and feature recognition open up novel avenues in the exploration for OAM beams, particularly in the recognition of OAM. Research in this field can be classified into two categories, classification and regression of photons' OAMs. Classification of photons' OAMs predicts the OAM values from a given list, while the regression of photons' OAMs predicts the OAM spectrum of an unknown state. In other words, classification of OAM states resolves the discretized OAM spectrum recognition task \cite{PhysRevA.104.053525,PhysRevLett.124.160401,Avramov-Zamurovic:23}; while regression of OAM states resolves the continuous OAM spectrum recognition task which requires more samples for training \cite{PhysRevA.103.063704,PhysRevApplied.17.054019,PhysRevResearch.5.013142}. 

There are various application of ML in OAM-based optical communication. For example, the ML technique has been applied in free-space optical communication to correct the deformation of the optical modes caused by turbulences \cite{Li:18,Liu:19,https://doi.org/10.1002/qute.202000103,na2021adaptive}. Also, the ML technique has assisted the recognition of fractional optical vortex modes \cite{PhysRevLett.123.183902,na2021deep,10.1063/5.0061365,PhysRevA.106.013519}, which opens a new way for high-capacity optical communication. In spite of the superhigh resolution of fractional optical vortex modes (98\% accuracy with OAM interval of 0.01 \cite{PhysRevLett.123.183902}), the fractional optical vortex modes are not ideal in practical application since they quickly diffract due to the phase discontinuity. Moreover, the distortion of the fractional optical vortex modes in the generating process would affect the accuracy apparently.

In this work, we apply the ML technique in the OAM state recognition for high-capacity optical communication. We discretize transverse spatial modes of structured light into individual partitions. Then we use the ML technique to effectively classify the OAM states. In this way, we turn the OAM state regression into classification, thus reducing the requirement of training samples. For the OAM states in each partition, we assign them a classical information. In other words, quantum information is approximately represented by classical information. We further apply our method to an image transmission task, demonstrating the ability to encode large data with low OAM number.

The key idea of this work is the discretization of the state space of light's spatial modes, which has not been reported before. The spatial modes of light are fully harnessed by considering the azimuthal and radial indexes of LG modes, which transmit stably in free space. In addition, since the OAM states in each partition represent the same information, our method allows for imperfection of OAM states in the generating process. Our method aims to overcome the challenges associated with the experimental realization of the OAM states, particularly the limitations on the orders of the OAM states.

\section{Methods}
\subsection{Dimension of the transverse spatial modes of structured light}
The transverse structure of paraxial beams propagating in free space can be described by different classes of orthonormal sets. The Hermite-Gaussian (HG) modes and the Laguerre-Gaussian (LG) modes are two widely-studied classes used to describe the transverse structure of a light beam. For a beam with wave number $k$ and propagating along $z$ axis, the HG modes and the LG modes have the following complex amplitudes \cite{BEIJERSBERGEN1993123}
\begin{eqnarray}
&&HG_{n,m}=C_1\frac{1}{w}e^{-ik\frac{(x^2+y^2)}{2R}}e^{-\frac{(x^2+y^2)}{w^2}}e^{-i(n+m+1)\Psi}H_n(\frac{x\sqrt{2}}{w})H_m(\frac{y\sqrt{2}}{w}),\\
\nonumber &&LG_{n,m}=C_2\frac{1}{w}e^{-ik\frac{(x^2+y^2)}{2R}}e^{-\frac{(x^2+y^2)}{w^2}}e^{-i(n+m+1)\Psi}(-1)^{\text{min}(n,m)}(\frac{r\sqrt{2}}{w})^{|n-m|}L_{\text{min}(n,m)}^{|n-m|}(\frac{2r^2}{w^2}),\\
\end{eqnarray}
where $C_1$ and $C_2$ are normalized constants, $R(z)=(z_r^2+z^2)/z$, $w(z)=\sqrt{2(z_r^2+z^2)/(kz_r)}$ and $\Psi=\arctan(z/z_r)$. $z_r$ is the Rayleigh length, $H_{n(m)}(\cdot)$ is the $n\text{th}$ $(m\text{th})$ order Hermite polynomial, and $L_p^{\ell}(\cdot)$ is the generalized Laguerre polynomial. The radial index $p$ and the azimuthal index (i.e., the OAM number) $\ell$ of LG modes are $p=\text{min}(n,m)$ and $\ell=n-m$, respectively. In this paper, we use the notation $LG_{n,m}$ to denote LG modes in accordance with the HG modes for order $N=n+m$, while we use the notation $LG(\ell,p)$ to explicitly denote the LG modes with OAM values $\ell$ and radial number $p$. The LG modes and the HG modes can be transformed through unitary operations \cite{BEIJERSBERGEN1993123}
\begin{equation}
\label{transform}
LG_{n,m}=\sum_{k=0}^{N}i^kb(n,m,k)HG_{N-k,k},
\end{equation}
where 
\begin{equation}
b(n,m,k)=\left(\frac{(N-k)!k!}{2^Nn!m!}\right)^{1/2}\times\frac{1}{k!}\frac{\text{d}^k}{\text{d}t^k}[(1-t)^n(1+t)^m]_{t=0},
\end{equation}
and $N=n+m$ is the order of the optical mode.

One can see from Eq.(\ref{transform}) that an LG beam can be decomposed into superposition of $N+1$ HG beams with the same order, and vice versa. Therefore, one could define the dimension of a subspace of the transverse mode $D=N+1$. In a subspace of dimension $D$, the LG modes and the HG modes could be transformed through unitary operations. For example, the LG$(\ell, p)$ modes LG(2,0), LG(-2,0) and LG(0,2) constitute a set of basis in a 3-dimensional subspace, while the HG$(m,n)$ modes HG(1,1), HG(2,0), and HG(0,2) constitute another set of basis. 

Within the scope of OAM recognition research, most existing studies primarily set the radial index $p$ to zero. While this simplification eases analysis, it overlooks the complexity arising from the combination of radial and azimuthal indexes. Here, we consider the complete spatial modes of structured light by taking account for the index $p$. By harnessing the complex spatial modes of structured light, we obtain a higher dimension of the transverse modes than only utilizing the OAM index $\ell$.

\subsection{Discretization of the state space}
\begin{figure}[htbp]
\centering\includegraphics[width=0.45\textwidth]{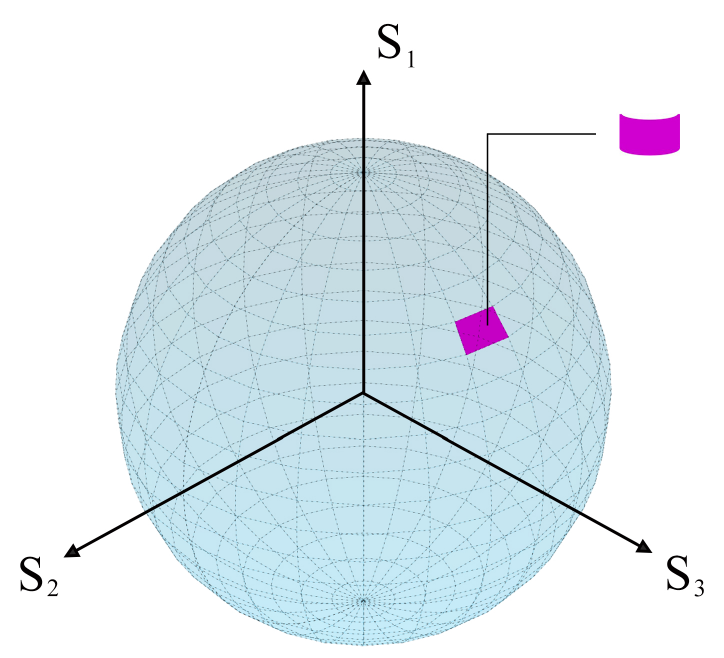} 
\caption{Discretization of the Bloch sphere. The polar angle and the azimuthal angle are uniformly divided, thus the surface of the Bloch sphere is partitioned into several regions. Each region is endowed with a classical state.}
\label{fig1}
\end{figure}

A quantum state in an $N$-dimensional quantum space can be represented by a state vector of $N$ elements. In quantum information, a $d$-state quantum system ($d>2$) is called a qudit, while a 2-state quantum system is called a qubit. For a qubit, an arbitrary pure quantum state can be represented by a point on the surface of the Bloch sphere. Here, we divide the spherical surface into several regions, each of which contains many states. We assign each region with a classical state, i.e., all the states in the region represent the same classical information. In other words, a qubit is approximated to $n$ classical bits, with $2^n$ being the number of the partitions of the spherical surface. 

A geometrical representation of such process is shown in Fig. \ref{fig1}. A general qubit state has the form
\begin{eqnarray}
|\phi\rangle=\cos\frac{\theta}{2}|0\rangle+\sin\frac{\theta}{2}e^{i\varphi}|1\rangle,
\end{eqnarray}
where $\theta$ and $\varphi$ are the polar angle and the azimuthal angle of the sphere. For example, we divide $\theta$ and $\varphi$ into 8 regions, respectively. The spherical surface is thus discretized into 64 regions. Therefore, a qubit is approximated to 6 classical bits. For a higher dimensional quantum space, one can follow the same procedure to discretize the space into several regions. 

The proposed discretization process in this work turns the OAM states regression into OAM states classification. Note that when the partitions of $\theta$ and $\varphi$ go to infinity, the problem becomes dividing OAM states into infinite categories, which is equivalent to OAM state regression. 

\subsection{Deep learning model}
\begin{figure}[htbp]
\centering\includegraphics[width=\textwidth]{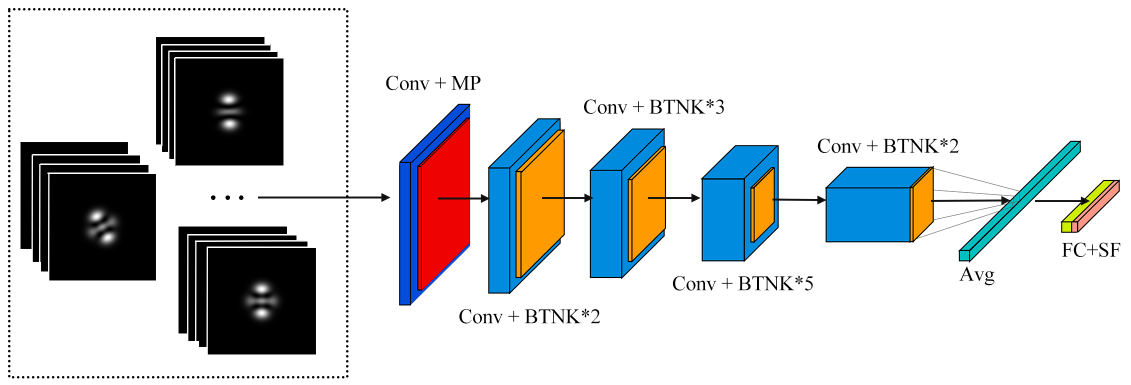} 
\caption{Neural network model for OAM-Recognition. The model architecture is based on an enhanced version of ResNet50, featuring a dual-stage structure for training. It includes convolutional layers (Conv) for feature extraction, batch normalization (BTNK) for stable training, max pooling (MP) for spatial reduction, fully connected Layers (FC) for complex pattern learning, and softmax activation (SF) for multi-class classification.}
\label{nnw}
\end{figure}

We develop an OAM-recognition neural network using deep learning technique shown in Fig. \ref{nnw}, which is pivotal for managing intricate tasks like image recognition and classification \cite{EGMONTPETERSEN20022279,PhysRevLett.123.183902}. Our study primarily concentrates on an enhanced version of a distinguished member of the Residual Network (ResNet) family \cite{Targ2016ResnetIR}, the ResNet50. This version integrates the concept of residual blocks and skips connections, effectively tackling the issue of vanishing gradients often encountered during complex image training. Each residual block comprises a sequence of convolutional layers, particularly the BottleneckBlock units \cite{Srinivas2021BottleneckTF}. These units start with a $1\times1$ convolution to reduce data dimensions, followed by a $3\times3$ convolution for a comprehensive feature extraction, and conclude with another $1\times1$ convolution to restore the data dimensions. Moreover, our discretization model adopts the CrossEntropyLoss function, which combines log softmax and negative log-likelihood loss to accurately compute gradients. We have implemented the ReduceLROnPlateau scheduler for dynamic adjustment of the learning rate, adapting it based on trends in validation loss to mitigate overfitting risks. Additionally, our model incorporates data augmentation and dropout techniques to enhance its generalization across various OAM states.

Regarding the model's architecture, we have devised a novel two-stage structure specifically tailored for the systematic processing of optical OAM information. In the initial stage, the model conducts a broad categorization of fundamental layers, distinguishing between various OAM categories presented in optical images. The subsequent stage involves a detailed sub-classification based on the primary layer's categorization, focusing on subtle variations in phase weights within optical images. By leveraging the inherent advantages of existing datasets, our model divides the OAM-recognition classification task into two stages, effectively overcoming challenges commonly associated with complex and high-dimensional datasets. This approach achieves rapid and high-precision classification while also partially reducing computational costs.

\section{Results}
\begin{figure}[htbp]
\centering\includegraphics[width=\textwidth]{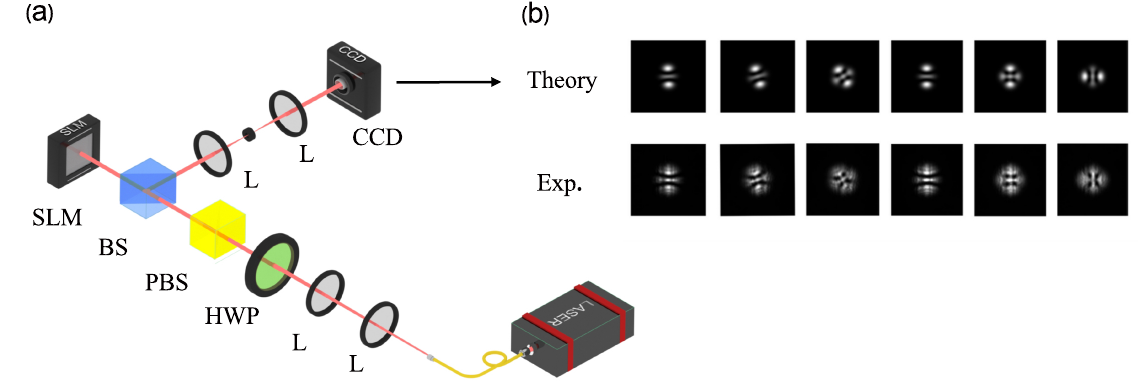} 
\caption{(a) Sketch of the experimental setup. (b) Examples of the generated OAM states in a three-dimensional space for six categories. SLM: spatial light modulator, HWP: half-wave plate, PBS: polarizing beam splitter, BS: beam splitter, L: lens.}
\label{exp}
\end{figure}

The experimental setup is shown in Fig. \ref{exp}. A coherent beam emitting from a laser diode is coupled into a single mode fiber. Then the beam is expanded by a $4f$ lens system to fit the size of the spatial light modulator (SLM). The SLM is loaded with gratings which generate the desired optical fields. The reflected beam is filtered by an aperture to pick out the first-order diffraction beam which is finally recorded by a CCD camera. Examples of experimentally generated structured light in a three-dimensional space are shown in Fig. \ref{exp}(b).

\begin{figure}[htbp]
\centering\includegraphics[width=0.5\textwidth]{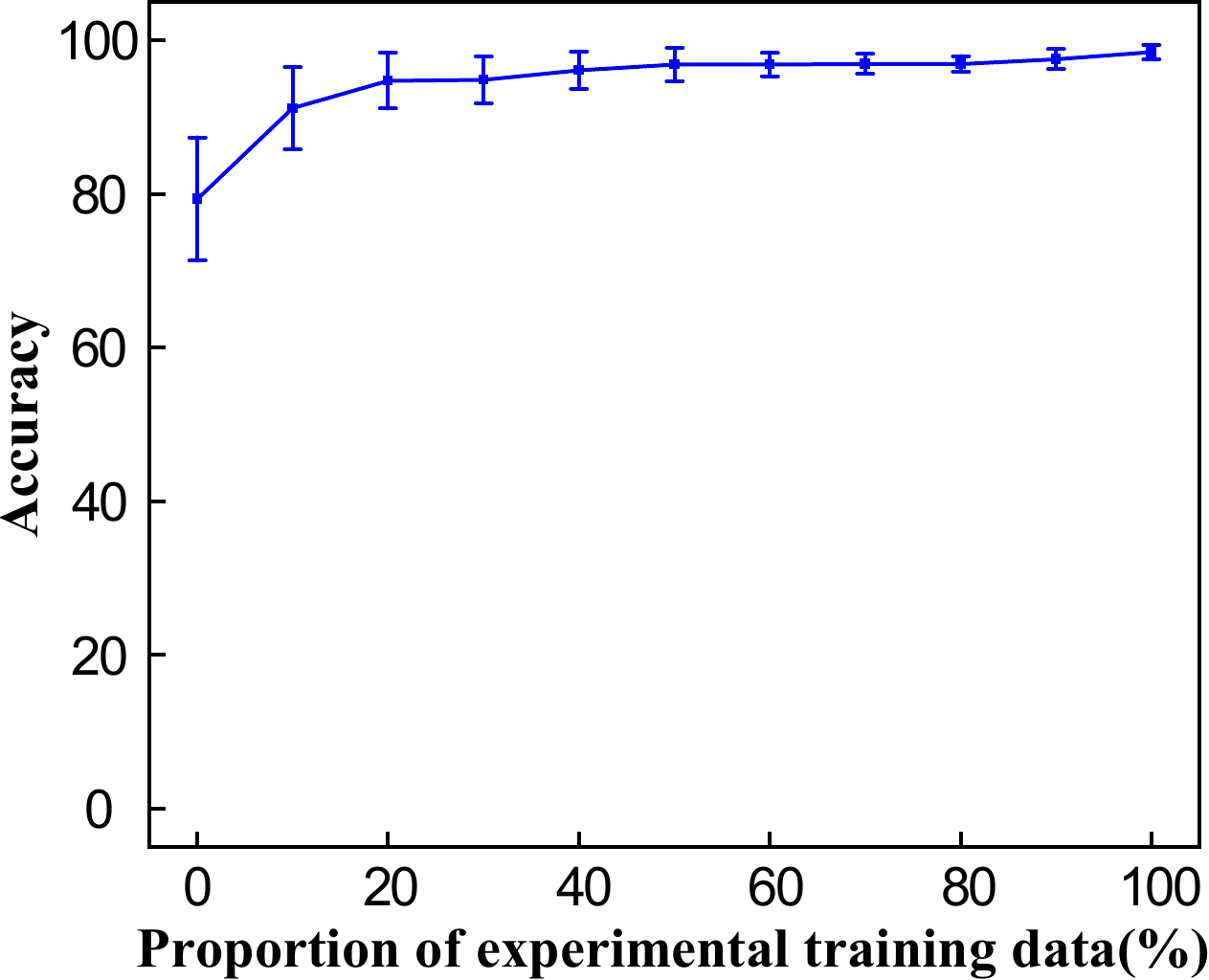} 
\caption{Prediction accuracy versus proportion of experimental images in the training set.}
\label{n31}
\end{figure}

The recorded images are then used for training. For a three-dimensional space, the basis states are LG(1,0), LG(-1,0) and LG(0,1). A general quantum state in such a space has the following form
\begin{eqnarray}
|\psi\rangle=\alpha|LG(1,0)\rangle+\beta|LG(-1,0)\rangle+\gamma|LG(1,1)\rangle,
\end{eqnarray}
where $\alpha, \beta$ and $\gamma$ are complex numbers which are normalized as $|\alpha|^2+|\beta|^2+|\gamma|^2=1$. In our experiment, we focus on the states with equal-weighted superposition, i.e., $\alpha=1/\sqrt{3}$, $\beta=e^{i\varphi_1}1/\sqrt{3}$, $\gamma=e^{i\varphi_2}1/\sqrt{3}$. Due to the degeneracy of the intensity profile of the OAM states $|\pm\ell\rangle$, we confine $\varphi_1$ and $\varphi_2$ between 0 and $\pi$. We further divide $\varphi_1$ and $\varphi_2$ into 30 intervals. We fix $\varphi_1=0$ and change $\varphi_2$, then fix $\varphi_2=0$ and change $\varphi_1$. In this way, we obtain 60 categories for a three-dimensional space. For each category, we generate 100 random samples for the model training, which are segmented into a training set (80 images ) and a validation set (20 images). For instance, for the first category with $\varphi_1=0$ and $\varphi_2\in(0,\pi/30)$, the samples are obtained by loading holograms which generate the state $|\psi\rangle$ with $\varphi_2$ uniformly distributed between 0 and $\pi/30$. We capture images of these states using a CCD camera and label them based on their OAM parameters. It is worth noting that one can extend the phase to the range $[0, 2\pi]$, with additional effort to break the degeneracy of the intensity profile of the states $|\pm\ell\rangle$ by taking one more picture \cite{PhysRevA.103.063704}.

\begin{figure}[htbp]
\centering\includegraphics[width=\textwidth]{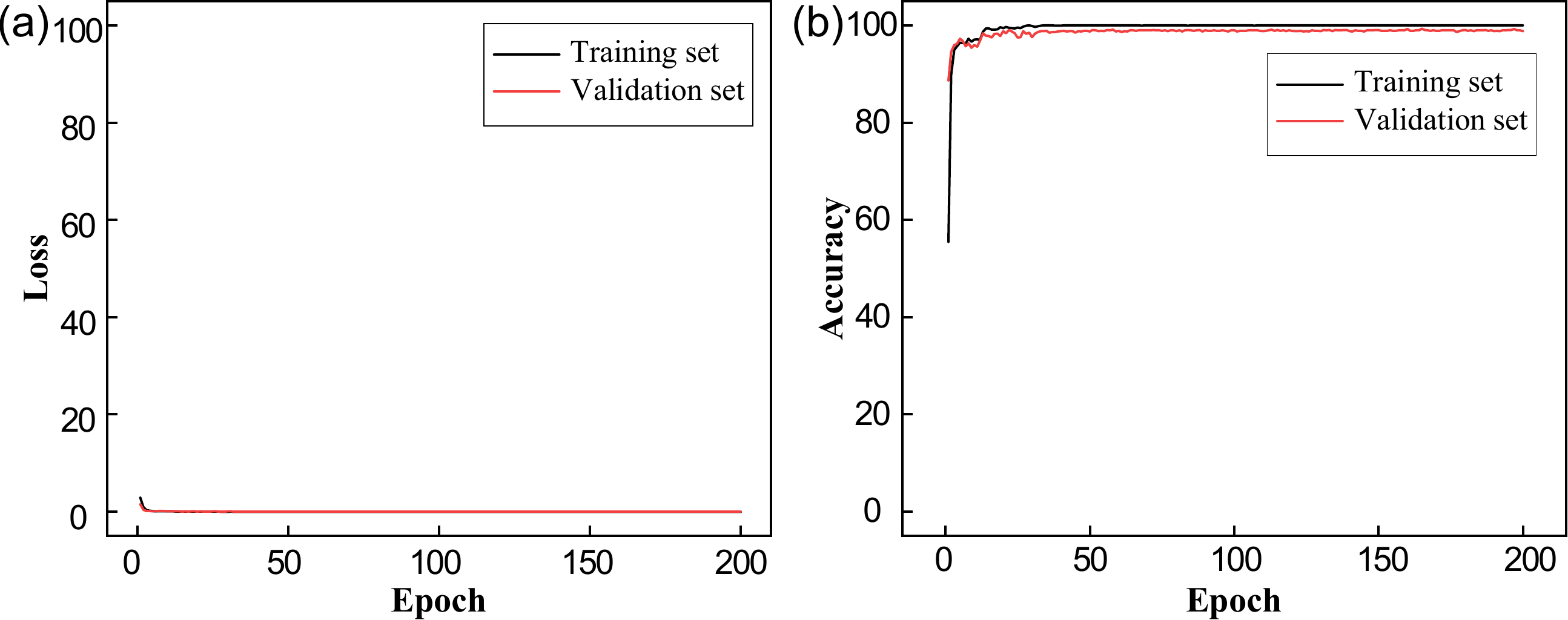} 
\caption{The loss (a) and the accuracy (b) in each epoch for the training and validation sets in the case $D=3$.}
\label{n3}
\end{figure}

\begin{figure}[htbp]
\centering\includegraphics[width=1\textwidth]{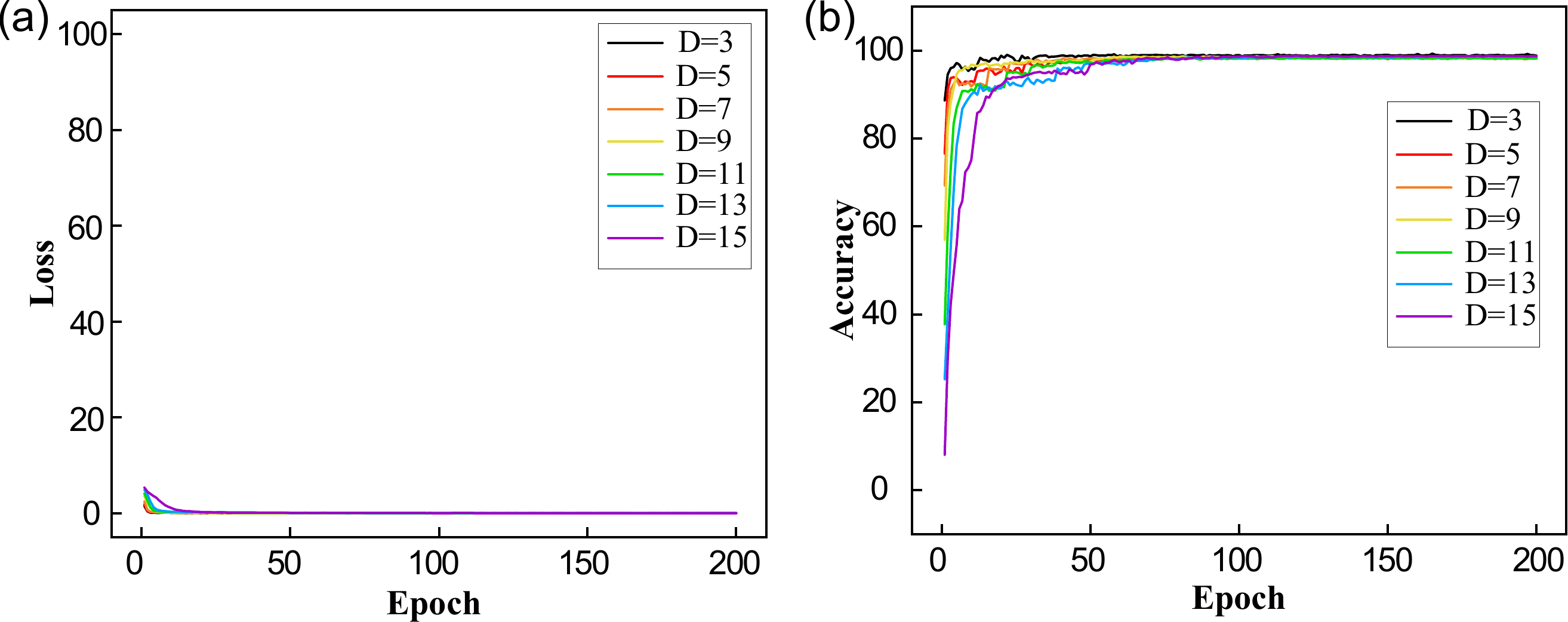} 
\caption{The loss (a) and the accuracy (b) in each epoch for the validation sets at different dimensions.}
\label{resul}
\end{figure}

We first train the model using a theoretical dataset. For the validation process, we use a test set entirely composed of experimental data. We only achieve a relatively low validation accuracy of $80\%\pm 10\%$. Moreover, incorporating experimental images into the training set allows for a more accurate simulation of actual experimental conditions. Therefore, to enhance the model's robustness, we gradually increase the proportion of experimental images in the training set to  20$\%$, 40$\%$, 60$\%$, and 80$\%$, while the validation set consistently comprises of the experimental images only. The result is shown in Fig. \ref{n31}. This incremental integration approach leads to a continuous improvement in accuracy, culminating in an optimal performance of 99.33$\%$ accuracy when both training and validation sets consist purely of the experimental data. Figure \ref{n3} shows the training evolution through the epochs. The loss and the accuracy of the validation set and the training set both show the consistent trend across epochs. The model obtains a relatively stable performance after about epoch 20.



We further apply our model for different dimensions, $D=5,7,9,11,13,15$. For each case, we discretize the state space by dividing the relative phases of the base states between 0 and $\pi$ into 30 partitions. Therefore, for the $D$-dimensional case, there are $30(D-1)$ categories. The results are shown in Fig. \ref{resul}. The loss and the accuracy of the validation set converge to stable values after about epoch 50. It is worth noting that our approach also allows for stacked classifications across different dimensions, accommodating up to 1680 distinct classifications for all seven dimensions, from $D=3$ to $D=15$. The accuracies for different dimensions are summarized in Table \ref{tab}. It is evident that the accuracy does not depend on the dimensions. As the dimensions grows, the accuracy maintains high. Therefore, our method can be generalized to the case of high-dimension space.

\begin{table}[htbp]
\centering
\caption{Accuracies for different dimensions.}
\begin{tabular}{cccccccc}
\hline
Dimension & 3 & 5 & 7 & 9 & 11 & 13 & 15 \\
\hline
Accuracy & 99.33\% & 99.23\% & 99.02\% & 99.18\% & 99.05\% & 99.15\% & 99.25\%\\
\hline
\end{tabular}
  \label{tab}
\end{table}

\begin{figure}[htbp]
\centering\includegraphics[width=\textwidth]{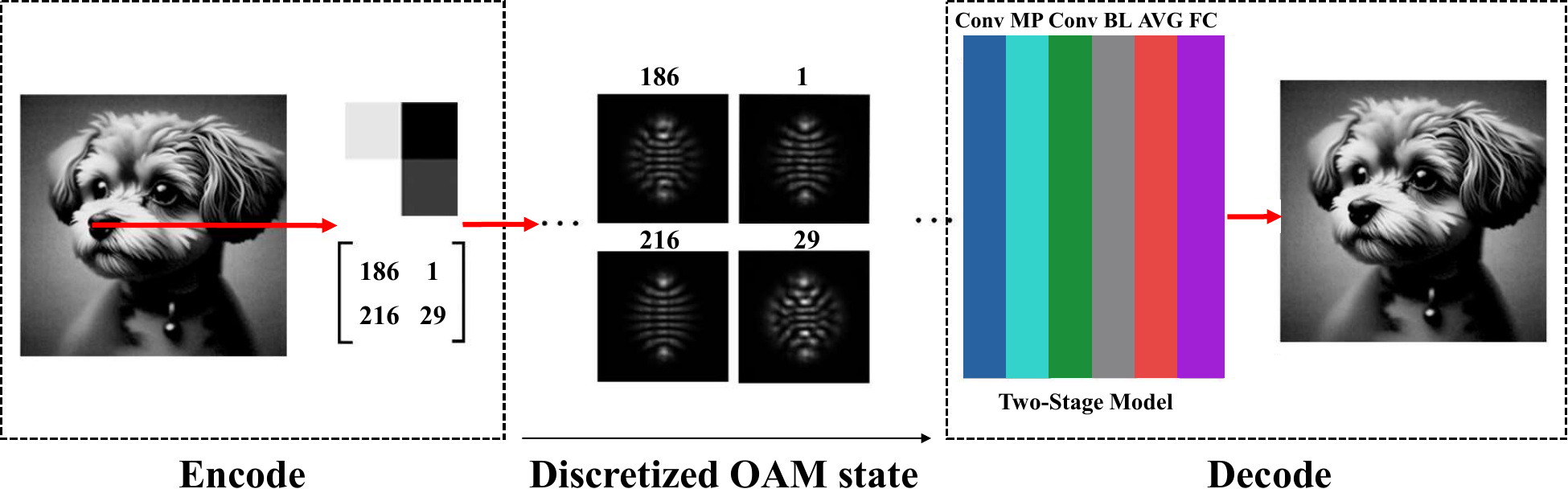} 
\caption{Procedure of an image transmission task. We map each pixel value of the image to the corresponding category of OAM states, and feed this data into our NNW model for prediction.}
\label{dog}
\end{figure}

To validate the applicability of our model, we apply the NNW model to an image transmission task. We aim to transmit a gray image of resolution $100\times 100$. We utilize the OAM states in a nine-dimensional space, thus there are 240 categories in such a space. To map the categories onto the pixel values of the image, the pixel values of the image are normalized between 0 and 239. Each pixel value corresponds to a category associated with a specific phase interval. The procedure is illustrated in Fig. \ref{dog}. The model exhibits exceptional ability in accurately predicting the specific OAM category for each pixel. The predicted category achieves high accuracy, underscoring its powerful predictive capabilities. The result highlights the efficiency of our model in decoding and reconstructing images based on the OAM categorized data. This comparison reveals a high degree of visual similarity, confirming the accuracy of our model and its practicality in optical communication.

\section{Conclusion}
We have developed a method for OAM state recognition, by discretizing the state space of the transverse spatial modes of structured light and specifically designing a tailored neural network architecture. Different from previous works, we consider the combination of the azimuthal and radial indexes of LG modes, which fully describe the transverse spatial modes of structured light. As a result, the LG modes transmit stably in free space. For the OAM states in each category, we assign them with the same classical information. Therefore, our method is robust to the deviation in the generating process of the spatial modes. The constructed neural network model, based on an enhanced ResNet50 architecture, achieves an accuracy of 99.33\% to classify the OAM states for the three-dimensional case and an accuracy of $99.25\%$ for the fifteen-dimensional case. The accuracy maintains a high value for different dimensions. Furthermore, our exploration of image transmission using this model with low OAM number has also yielded positive outcomes, highlighting its practical value in high-capacity optical communication applications. These advances open up new avenues for information transmission and contribute significantly to the field of optical communication and data encoding technology.


\section*{Funding}
This work is supported by the National Natural Science Foundation of China (Grant Nos. 12204312, U21A20436, 12104190), the Jiangxi Provincial Natural Science Foundation (Grant Nos. 20224BAB211014 and 20232BAB201042), the Natural Science Foundation of Jiangsu Province (Grant No. BK20210874), and the General Project of Natural Science Research in Colleges and Universities of Jiangsu Province (Grant No. 20KJB140008).


\section*{Disclosures}
The authors declare no conflicts of interest.

\section*{Date availability}
Data underlying the results presented in this paper are not publicly available at this time but may be obtained from the authors upon reasonable request.



\end{document}